\documentstyle[prd,aps,preprint,tighten,]{revtex}

\newcommand{\be} {\begin{equation}}
\newcommand{\ee} {\end{equation}}
\newcommand{\beann}{\begin{eqnarray*}}
\newcommand{\eeann}{\end{eqnarray*}}
\newcommand{\bea}{\begin{eqnarray}}
\newcommand{\eea}{\end{eqnarray}}
\newcommand{\nn}{\nonumber}

\newtheorem{thm}{Theorem}

\begin{document} 

\draft

\title{Black hole interacting with matter as a simple dynamical system}  
  
\author{Petr H\'{a}j\'{\i}\v{c}ek} 
\address{Institute for Theoretical Physics\\
   University of Bern\\
   Sidlerstrasse 5, CH-3012 Bern\\
   Switzerland}

\date{\today} 
\maketitle

\begin{abstract} Recently, a variational principle has been derived
  from Einstein-Hilbert and a matter Lagrangian for the spherically symmetric
  system of a dust shell and a black hole. The so-called physical region of
  the phase space, which contains all physically meaningful states of the
  system defined by the variational principle, is specified; it has a
  complicated boundary. The principle is then transformed to new variables
  that remove some problems of the original formalism: the whole phase space
  is covered (in particular, the variables are regular at all horizons), the
  constraint has a polynomial form, and the constraint equation is uniquely
  solvable for two of the three conserved momenta. The solutions for the
  momenta are written down explicitly. The symmetry group of the system is
  studied. The equations of motion are derived from the transformed principle
  and are shown to be equivalent to the previous ones. Some lower-dimensional
  systems are constructed by exclusion of cyclic variables, and some of their
  properties are found.
\end{abstract} 


\section{Introduction}
The spherically symmetric gravitating shell has been used as a simplified
quantum model at many occasions. One can mention the quantum effects of domain
walls in the early Universe \cite{guth}, quantum aspects of gravitational
collapse \cite{HKK}, quantum theory of black holes \cite{hooft} or Hawking
evaporation of black holes \cite{KW}.

The equations of motion for the shells has been derived by Dautcourt \cite{D}
and transformed to a geometric (i.e.\ ``gauge invariant'') form by Israel
\cite{I}. However, for a quantum theory, one needs a variational principle
rather than dynamical equations. Such variational principle has as yet been
just guessed from the dynamical equations, Refs.\ \cite{berezin} and
\cite{HB}, or from some intermediate variational principle, Refs.\ \cite{KW}
and \cite{louko}. Quite a number of different Hamiltonians have resulted and
the corresponding quantum theories have not been unitarily equivalent. There
are two sources of the ambiguity.

The first source of ambiguity is the invariance of the general relativity with
respect to any coordinate transformations on one hand, and the property of
Hamiltonians to generate dynamics with respect to a particular time coordinate
on the other. For example, the choice of the proper time along the shell as
such a coordinate leads to the cosh$p$ Hamiltonian, Refs.\ \cite{berezin} and
\cite{H3}, the Schwarzschild time coordinate inside the shell to the
square-root Hamiltonian, Ref.\ \cite{HKK}, and the Schwarzschild time
coordinate outside the shell to a merely implicitly determined Hamiltonian of
Refs.\ \cite{KW} and \cite{louko}. An attempt to work out a gauge invariant
theory for at least some class of the guessed Hamiltonians is Ref.\ \cite{HB};
based on a technical assumption that the super-Hamiltonian is quadratic in the
momenta, it gives a unique action principle.

The second source of ambiguity is the fact that equations of motion do not
determine the corresponding variational principle uniquely in general. This
ambiguity can be removed by a direct derivation of the variational principle
for the shell from the Einstein-Hilbert and shell matter Lagrangian. Three
such derivations exist: Refs.\ \cite{Hkij}, \cite{ansold} and \cite{Khaj};
they all lead to the same variational principle for the same system. In Refs.\ 
\cite{ansold} and \cite{Khaj}, the reduction of the second order formalism to
the spherically symmetric case is done first, followed by a reduction to
dynamical variables of the shell alone and by the transformation to the second
order formalism. In Ref.\ \cite{Hkij}, a first and second order formalism for
a general shell and gravitational field is derived (no symmetry); this second
order formalism is reduced by the spherical symmetry and to shell variables in
Ref.\ \cite{H1}.

We are then left with the following dilemma. The super-Hamiltonian of Ref.\ 
\cite{HB} is simple and amenable to the existing gauge invariant quantization
methods (like e.g.,\ Refs.\ \cite{HK1} and \cite{HK2}). The super-Hamiltonian
of Refs.\ \cite{ansold}, \cite{H1} and \cite{Khaj} follows from the
Einstein-Hilbert and shell-matter Lagrangian, but it is extremely complicated:
it contains nested square roots and hyperbolic functions, and for a good
measure, it is formulated in coordinates that diverge on horizons. It would be
difficult to quantize.

A natural question then arise: what is the relation between the guessed and
simple variational principle of Ref.\ \cite{HB} and the derived but
complicated one of Refs. \cite{ansold}, \cite{H1} and \cite{Khaj}? If the
simple principle were equivalent to the complicated one, we could use the
simple principle and forget the complicated one. However, the answer turns out
to be rather surprising (see Ref.\ \cite{H2}): the two descriptions are only
locally, but not globally equivalent. The local equivalence explains, why the
equations of motion are the same. For the study of the problem, a geometric
approach of Ref.\ \cite{nucl} has been applied. All gauge invariant properties
of a constrained system are encoded in two objects: the constraint manifold
$\Gamma$ and the presymplectic form $\Omega$ on $\Gamma$.  The equivalence of
two such systems, $(\Gamma_1,\Omega_1)$ and $(\Gamma_2,\Omega_2)$ is then a
well-defined mathematical concept independent of a choice of extended phase
space, a choice of constraint functions or a choice of gauge.

In this situation, it seems natural to look for a transformation of the
variational principle of Refs.\ \cite{ansold}, \cite{H1} and \cite{Khaj} to a
better set of coordinates, and this will be the main topic of the present
paper. The method will again be based on the gauge invariant description
$(\Gamma,\Omega)$ of reparametrization invariant systems.  We shall perform
the transformation in three steps; to motivate the steps, we take into account
the symmetry of the system, the structure of the so-called {\em Cartan form}
$\Theta$, which is defined by $d\Theta := \Omega$, and the topology of the
space $(\Gamma,\Omega)$. The topological problem involved here is the
existence of transversal surfaces in $\Gamma$; such surfaces are nowhere
tangential to the dynamical trajectories. Functions whose levels are
transversal are not only helpful, if one looks for nice coordinates, because
they simplify $\Theta$. Quite generally, they have to do with the existence of
Hamiltonian and with the possibility to give the quantum dynamics the form of
Schr\"{o}dinger equation (cf.\ Ref.\ \cite{nucl}). We shall, therefore, also
look for transversal surfaces in a more systematic way. Finally, we shall find
that the structure of the phase space of the spherically symmetric shell is
not simple. Only a proper subset, which we call {\em physical region},
contains physical states of the shell. Its nature can be roughly specified as
follows.  All points inside it correspond to the system consisting of a shell
of positive mass and radius, interacting with a black hole. The boundary of
the region is complicated; it contains shells with zero radius, shells with
zero rest mass (which cannot, however, be regarded as shells made of
light-like matter), as well as self-gravitating isolated shells with positive
radius and mass---these are the points at the boundary that are physically
meaningful.  The points outside the region that describe shells with positive
radius and mass correspond to systems consisting of the shell and a negative
mass source.  A conclusion seems to be that the system will be difficult to
quantize even if the new variables has made it algebraically simple. One
problem will be to keep the spectra of observables contained within the
physical region. Another problem will be to construct a unitary dynamics,
because the classical dynamics breaks down at the boundary of the physical
region.

The plan of the paper is as follows. In Sec.\ \ref{sec:descr}, we collect the
relevant results of the previous papers, mainly Refs.\ \cite{H1} and
\cite{Khaj}. The variational principle that we describe allows also for an
internal degree of freedom of the shell; it depends, therefore, of two
additional variables in comparison with Refs.\ \cite{ansold} and \cite{H1}:
the proper time along the shell trajectory and the rest mass of the shell. We
determine the constraint manifold $\Gamma$ and the Cartan form $\Theta$ that
follow from the variational principle. We study the physical region.

In Sec.\ \ref{sec:mom}, we introduce three new functions on the phase space
that are regular at the horizons and that will replace the momentum $P$ of the
shell and the two Schwarzschild times $T^\pm$ that diverge there. We describe
their physical and geometrical meaning, using also some results of previous
papers. In Sec.\ \ref{sec:coord}, we specify a new set of coordinates on
$\Gamma$, and show that the transformation from the old to the new coordinates
is differentiable and invertible. We also find the ranges of these coordinates
in the physical region. Sec.\ \ref{sec:symm} collects and extends the results
about symmetries of the system that were obtained in Ref.\ \cite{H2} and
expresses the symmetry transformations in the new coordinates; the form of the
transformation simplifies. The new times are cyclic coordinates as the old
have been.  In Sec.\ \ref{sec:cartan}, we transform the Cartan form into new
coordinates. The square roots and hyperbolic functions disappear, and an
equivalent form becomes regular at the horizons so that it can be smoothly
matched accross them. However, the Cartan form is still complicated and this
motivates further transformations. The final result thereof can be described
as follows.  The extended phase space is $({\mathbf R}^8,\bar{\Omega})$. The
natural coordinates of ${\mathbf R}^8$ form a Darboux chart for
$\bar{\Omega}$. The equation of the constraint surface $\Gamma$ in ${\mathbf
  R}^8$ is polynomial in all momenta and the presymplectic form $d\Theta$ is
the pull-back of $\bar{\Omega}$ to $\Gamma$.

In Sec.\ \ref{sec:constr}, we study the polynomial constraint. We show that it
defines two disjoint submanifolds in ${\mathbf R}^8$. One, denoted by
$\Gamma_6$, is six-dimensional and lies well outside the physical region. The
other, denoted by $\Gamma_7$, is seven-dimensional and intersects the physical
region. In the intersection of $\Gamma_7$ with the physical region, we find
two different foliations by transversal surfaces. They are not globally
transversal, however, because none of them is intersected by all dynamical
trajectories. The problem is that each dynamical trajectory starts or finishes
at the ($R=0$)-subset of the boundary of the physical region. The two
different Hamiltonians coresponding to the two foliation are written down
explicitly; they contain second and third roots.

In Sec.\ \ref{sec:cases}, we repeat the study for some important
lower-dimensional cases; they can be obtained by exclusion of cyclic
coordinates from the original system. In particular, a (dynamical) black hole
and the shell without its internal degree of freedom, the shell in the field
of a fixed black hole external field, or the isolated self-gravitating shell
(with the flat spacetime inside) are considered.

\section{Description of the system}
\label{sec:descr}
In this section, we shall collect some results of Refs.\ \cite{HB}, \cite{H1}
and \cite{Khaj} so that the paper becomes relatively self-consistent. We shall
also describe and study the physical region.

A spherically symmetric thin-shell spacetime solution of Einstein equations
can be constructed as follows.  Consider two Schwarzschild spacetimes
${\mathcal M}_1$ and ${\mathcal M}_2$ with Schwarzschild masses $E_1$ and
$E_2$. Let $\Sigma_1$ be a timelike hypersurface in ${\mathcal M}_1$ and
$\Sigma_2$ be one in ${\mathcal M}_2$.  Let $\Sigma_1$ divide ${\mathcal M}_1$
into two subspacetimes, ${\mathcal M}_{1+}$ and ${\mathcal M}_{1-}$, and
similarly $\Sigma_2$ divide ${\mathcal M}_2$ into ${\mathcal M}_{2+}$ and
${\mathcal M}_{2-}$. As everything is spherically symmetric, all spacetimes
are effectively two-dimensional; we chose fixed time and space orientation in
the two-dimensional Schwarzschild spacetimes so that future and past as well
as right and left are unambiguous; let then ${\mathcal M}_{2+}$ and ${\mathcal
  M}_{1+}$ be right with respect to ${\mathcal M}_{2-}$ and ${\mathcal
  M}_{1-}$. Let $\Sigma_1$ and $\Sigma_2$ be isometric; then the spacetime
${\mathcal M}_{1-}$ can be pasted together with the spacetime ${\mathcal
  M}_{2+}$ along the boundaries $\Sigma_1$ and $\Sigma_2$. The result is a
shell spacetime ${\mathcal M}_s$. Given a shell spacetime, we shall leave out
the indices 1 and 2, having right (left) energy $E_+$ ($E_-$), shell
trajectory $\Sigma$, and the right (left) subspacetime ${\mathcal M}_{+}$
(${\mathcal M}_{-}$). Thus, ${\mathcal M}_{+} = {\mathcal M}_{2} \cap
{\mathcal M}_{s}$ and ${\mathcal M}_{-} = {\mathcal M}_{1} \cap {\mathcal
  M}_{s}$. In Refs.\ \cite{H1}, \cite{H2} and \cite{HB}, the Schwarzschild
masses $E_\epsilon$, $\epsilon = \pm 1$ were assumed to be {\em positive}.

Each subspacetime ${\mathcal M}_\epsilon$ has the metric
\be
  ds_\epsilon^2 = - F_\epsilon(R)(dT^\epsilon)^2 + F_\epsilon^{-1}(R)dR^2, 
\label{Smetr}
\ee
where
\[
  F_\epsilon(R) := 1 - \frac{2E_\epsilon}{R},
\]
and it is split up by its horizons into four quadrants: we denote by $Q_I$
that which is adjacent to the right infinity, $Q_{II}$ to the left infinity,
$Q_{III}$ to the future singularity and $Q_{IV}$ to the past singularity.
Suitable notation (introduced in Ref.\ \cite{H1}) enables us to write all
formulas in a form valid in any quadrant: we define four sign functions
$a_\epsilon$ and $b_\epsilon$ distinguishing the quandrants of ${\mathcal
  M}_\epsilon$: $a_\epsilon := \text{sgn}F_\epsilon$, and $b_\epsilon$ equals
to $+1$ ($-1$) in the past (future) of the event horizon in the spacetime
$({\mathcal M}_\epsilon, g_\epsilon)$. We also include the dust internal
degree of freedom, $\sf M$, and the proper time $\sf T$ along the shell
trajectory using the results of Ref.\ \cite{Khaj}. Then the total action for
the dust and gravity can be rewritten in the form
\be 
  S_\Sigma[{\sf
  T},{\sf M};T^\pm,E_\pm;R,P;\nu] = \int dt(P\dot{R} - E_+\dot{T}^+
  + E_-\dot{T}^- + {\sf M}\dot{\sf T} - \nu{\mathcal C}), 
\label{act}
\ee
where $\nu$ is a Lagrange multiplier, 
\be
  {\mathcal C} = {\sf M} - R\sqrt{F_+ + F_- -
    2a_-b_+b_-\sqrt{|F_+F_-|}\text{sh}_{a_+a_-}\frac{P}{R}}, 
\label{supH}
\ee
is the super-Hamiltonian, $\text{sh}_ax := (\text{e}^x +a\text{e}^{-x})/2$ for
any $a = \pm 1$ and $x \in (-\infty,\infty)$. The momentum $P$ conjugate to
the radial coordinate $R$ can be defined as follows (for more detail, see
Ref.\ \cite{H1}). Let $(n_\epsilon,m_\epsilon)$ be the orthonormal dyad at a
shell point in the subspacetime $({\mathcal M}_\epsilon, g_\epsilon)$, the
vector $n_\epsilon$ being tangential to the shell and future oriented,
$m_\epsilon$ being right oriented; we call it {\em shell dyad}. Let
further the so-called {\em Schwarzschild dyad} $(n_{S\epsilon},m_{S\epsilon})$
be defined at each point of the shell in the spacetime $({\mathcal
  M}_\epsilon, g_\epsilon)$ by the requirement that $n_{S\epsilon}$ be
timelike future oriented, $m_{S\epsilon}$ be right oriented and that one of
the vectors be tangential to the ($R =$ const)-curves. Clearly, in each
quadrant, we have a different formula for the components of the Schwarzschild
dyad; with respect to the Schwarzschild coordinates,
\beann
  \text{for}\,F_\epsilon & > & 0,\quad n_{S\epsilon} =
  (b_\epsilon/\sqrt{F_\epsilon},0),\quad m_{S\epsilon} =
  (0,b_\epsilon\sqrt{F_\epsilon}); \\
  \text{for}\,F_\epsilon & < & 0,\quad n_{S\epsilon} =
  (0, b_\epsilon\sqrt{|F_\epsilon|}),\quad m_{S\epsilon} =
  (-b_\epsilon/\sqrt{|F_\epsilon|},0).
\eeann
Then, $P_\epsilon/R$ is the hyperbolic angle between the Schwarzschild and
shell dyads:
\beann
 n_\epsilon & = & n_{S\epsilon}\,\text{cosh}P_\epsilon/R +
 m_{S\epsilon}\,\text{sinh}P_\epsilon/R, \\
 m_\epsilon & = & n_{S\epsilon}\,\text{sinh}P_\epsilon/R +
 m_{S\epsilon}\,\text{cosh}P_\epsilon/R.
\eeann
Finally, $P:= [P]$, where we use the common short-hand for the jump $[A] :=
A_+ - A_-$ of a quantity $A$ across the shell.

The extended phase space of the system is eight-dimensional, split up into 16
disjoint sectors, each sector being a pair of quadrants, one chosen from the
left and one from the right subspacetimes. Each of these sectors can be
covered with the coordinates $P$, $R$, $E_+$, $T^+$, $E_-$, $T^-$, $\sf M$ and
$\sf T$. The constraint surface that we call $\Gamma$ is defined by the
constraint equation ${\mathcal C} = 0$ in the extended phase space. It does
not intersects all 16 sectors, but it is split up into at most (if $E_- > {\sf
  M}$) non-empty intersections of the sectors with it.\footnote{Observe that
  we admit unphysical sectors, i.e.\ those, in which the constraint equation
  ${\mathcal C} = 0$ has no solutions. This simplifies the boundary of the
  extended phase space a great deal and does not lead to any problem: it is
  the very purpose of extended phase spaces to enclose also non-physical
  poits, if this leads to any simplification.} These can be covered by the
system of seven coordinates $P$, $R$, $E_+$, $T^+$, $E_-$, $T^-$ and $\sf T$.
The following coordinates are assumed to have non-trivial ranges: 
\be 
  R > 0,\quad E_+ > 0,\quad E_- > 0,\quad {\sf M} > 0.
\label{ran1}
\ee

The pull back of the Liouville form of the action (\ref{act}) to the
constraint surface $\Gamma$ in these coordinates is simply
\be 
  \Theta = PdR - E_+dT^+ + E_-dT^- + {\sf M}d{\sf T},
\label{thet}
\ee
where $\sf M$ is now a function of the seven coordinates defined by
\be
  {\sf M} = R\sqrt{F_+ + F_- -
    2a_-b_+b_-\sqrt{|F_+F_-|}\text{sh}_{a_+a_-}\frac{P}{R}}. 
\label{constr}
\ee
The form (\ref{thet}) is called {\em Cartan form} of the system; it contains
all information about the dynamics and about the Poisson brackets (see, e.g.\ 
Refs.\ \cite{souriau} or \cite{H2}).

The present paper starts from these formulas and quantities. The first new
observation that we make is, that all these formulae remain valid also for the
values $E_+ = 0$ or $E_- = 0$, if the notation and conventions are adapted a
little. These cases are interesting because they describe an isolated
self-gravitating shell (with the flat spacetime inside). Indeed, the metric of
the flat spacetime is covered by formula (\ref{Smetr}). However, the flat
spacetime does not contain any horizon and is not split into quadrants.
Still, a piece of the flat spacetime can be used in four distinct ways
in the contruction of the shell spacetime: either it lies to the left
($E_-=0$) or to the right ($E_+=0$) of the shell, and, in each case, the $R$
coordinate can either increase to the right or to the left ($R$ is always
spacelike in the flat spacetime). We can, therefore, formally define the
$E$-{\em spacetime} for $E>0$ as before to be the Kruskal spacetime of Schwarzschild
mass $E$ with four quadrants, and for $E=0$ as consisting of two topologically
separated quadrants $Q_I$ and $Q_{II}$, each isometric to the flat spacetime,
but with oposite time and space orientations: in $Q_I$, the radial
(``Schwarzschild'') coordinate $R$ increases to the right and the
``Schwarzschild'' time $T$ to the future; in $Q_{II}$, $R$ increases to the
left and $T$ to the past. It is amusing to observe that the validity of the
formulae (\ref{act}) and (\ref{supH}) can then be extended to vanishing $E$'s,
if we just use the values $a_\epsilon=1$ everywhere, $b_\epsilon=+1$ in $Q_I$
and $b_\epsilon=-1$ in $Q_{II}$. This works because the components of the
Schwarzschild dyad that are crucial for the derivation of Eq.\ (\ref{supH}) in
Ref.\ \cite{H1} retain their form in the quadrants $Q_I$ and $Q_{II}$ of the
$E$-spacetime for $E=0$, and there are no quadrants $Q_{III}$ and $Q_{IV}$.
Similarly, the sign relation (14) of Ref.\ \cite{HB}, which is basic for the
form of the dynamical equations, remains valid with our orientation and
quadrant conventions. Observe that the presence of two topologically separated
subspacetimes is not observable and does not imply any physical assertion;
similarly, the orientation of coordinates is a mere coordinate convention
without any physical meaning.

A very important but rather embarrasing observation is now that a formally
impeccable shell spacetime can be pasted together from Schwarzschild
spacetimes of {\em any} mass, even a negative one. The metric is still given
by Eq.\ (\ref{Smetr}) and the spacetime has a similar global structure as the
flat one: there are no horizons and there are two possible orientations of how
it can be built in. Again, all equations remain valid, if we accept the same
sign, quadrant and orientation conventions as for the flat spacetime and
define the (two-quadrant) $E$-spacetime for $E<0$ analogously to what we have
done for $E=0$. The corresponding shell dynamics is then again given by the
action (\ref{act}).

Let us quickly discuss the dynamical equations following from the
action (\ref{act}) under these conventions. There are three conservation laws,
\be
  \dot{E}_+ = \dot{E}_- = \dot{\sf M} = 0,
\label{cons}
\ee
resulting from varying the action with respect to the cyclic
coordinates $T^+$, $T^-$ and $\sf T$. The variations with respect of the
conserved momenta and $P$ with the subsequent simplification by the constraint
(\ref{constr}) yield:
\bea
  \dot{\sf T} & = & \nu,
\label{dotTm} \\
  \dot{T}^\epsilon & = & -\epsilon\nu\frac{R}{\sf M}
  \left(1-a_-b_+b_-\frac{\sqrt{|F_+F_-|}}{F_\epsilon}\,
  \text{sh}_{a_+a_-}\frac{P}{R}\right),  
\label{dotTe} \\
  \dot{R} & = & \nu a_-b_+b_-\frac{R}{\sf M}\sqrt{|F_+F_-|}\,
  \text{sh}_{-a_+a_-}\frac{P}{R}.
\label{dotR}
\eea
The equation determining $\dot{P}$ can be obtained either by varying the
action with respect to $R$, or by differentiating the constraint equation
${\mathcal C} = 0$ with respect to $t$ followed by substitution for all other
$t$-derivatives from Eqs.\ (\ref{cons})--(\ref{dotR}). Let us rewrite the
constraint equation in the form
\be
  -\frac{\sf M}{2R} + \frac{R}{2{\sf M}}(F_+ + F_-) = a_-b_+b_-\frac{R}{\sf
    M}\sqrt{|F_+F_-|}\,\text{sh}_{a_+a_-}\frac{P}{R}.
\label{constr2}
\ee
Eqs.\ (\ref{cons})--(\ref{constr2}) represent a complete system of dynamical
equations for the shell. 

We derive some important consequences from the dynamical equations.
Eqs.\ (\ref{dotTm})--(\ref{constr2}) imply that
\[
  -F_\epsilon\left(\frac{dT^\epsilon}{d{\sf T}}\right)^2 +
  \frac{1}{F_\epsilon}\left(\frac{dR}{d{\sf T}}\right)^2 = -1,
\]
confirming that  $\sf T$ is the proper time along shell trajectories. 
Eqs.\ (\ref{dotTm}), (\ref{dotTe}) and (\ref{constr2}) deliver the {\em time
equation}: 
\be
  F_\epsilon\frac{dT^\epsilon}{d{\sf T}} = -\epsilon\frac{\sf M}{2R} +
  \frac{E_+-E_-}{\sf M}.
\label{timeeq}
\ee
From Eqs.\ (\ref{dotR}) and (\ref{constr2}), the {\em radial equation} follows,
\be
  \frac{dR}{d{\sf T}} = -\omega\sqrt{-V},
\label{radeq}
\ee
where
\be
  V(R) = -\frac{{\sf M}^2}{4R^2} - \frac{E_+ +E_-}{R} - \frac{(E_+
  -E_-)^2}{{\sf M}^2} + 1
\label{radpot}
\ee
and $\omega= \pm 1$ is a suitable sign (see Eq.\ (2) of Ref.\ \cite{H2}). The
square of the radial equation can be rewritten in one of the two forms
\be
  F_\epsilon + \left(\frac{dR}{d{\sf T}}\right)^2 = J^2_\epsilon,
\label{radeqe}
\ee
according as $\epsilon = \pm 1$, where
\be
  J_\epsilon := \frac{\sf M}{2R} - \epsilon\frac{E_+-E_-}{\sf M}
\label{J}
\ee
(cf.\ Ref.\ \cite{HB}). It follows immediately from this definition of
$J_\epsilon$ that
\[
  J_+ + J_- = \frac{\sf M}{R},
\]
and this can be transformed with the help of Eq.\ (\ref{radeqe}) to
\[
  \text{sgn}J_+\sqrt{F_+ + \left(\frac{dR}{d{\sf T}}\right)^2} +
  \text{sgn}J_-\sqrt{F_- + \left(\frac{dR}{d{\sf T}}\right)^2} = \frac{\sf
  M}{R}. 
\]
However, comparing Eqs.\ (\ref{timeeq}) and (\ref{J}), we observe that
\[
  \text{sgn}J_\epsilon =
  -\epsilon\,\text{sgn}\left(F_\epsilon\frac{dT^\epsilon}{d{\sf T}}\right).
\]
The sign of the expression $F_\epsilon(dT^\epsilon/d{\sf T})$ deserves a
name; let us call it $\tau_\epsilon$ (cf.\ \cite{HB}). Then, we obtain finally:
\be
  -\tau_+\sqrt{F_+ + \left(\frac{dR}{d{\sf T}}\right)^2} + \tau_-\sqrt{F_- +
    \left(\frac{dR}{d{\sf T}}\right)^2} = \frac{\sf M}{R}.
\label{Israel}
\ee
This is the so-called {\em Israel equation} for dust shell (cf.\ Eqs. (14),
(21) and (22) of Ref.\ \cite{HB}).

Eqs. (\ref{timeeq})--(\ref{Israel}) are valid for all values of Schwarzschild
masses $E_+ \in (-\infty,+\infty)$ and $E_- \in (-\infty,+\infty)$. Some
features of the unphysical (negative $E_\epsilon$) cases are described by the
following theorem.
\begin{thm}
  Let $E_\epsilon$ be non-positive and not larger than $E_{-\epsilon}$ for
  some value of $\epsilon$. Then the timelike (connected) center ($R=0$) curve
  lies in the subspacetime ${\mathcal M}_\epsilon$ on the $\epsilon$-side of
  the shell.
\end{thm}
To prove the theorem, we observe that $E_\epsilon \leq E_{-\epsilon}$ implies
\[
  \sqrt{F_\epsilon + \left(\frac{dR}{d{\sf T}}\right)^2} \geq
    \sqrt{F_{-\epsilon} + \left(\frac{dR}{d{\sf T}}\right)^2}.
\]
Thus, the left-hand side of Eq.\ (\ref{Israel}) can only be positive, if
$\tau_\epsilon = -\epsilon$. The assumption $E_\epsilon \leq 0$ of the theorem
implies that ${\mathcal M}_\epsilon$ is the two-quadrant $E$-spacetime with
$\tau_\epsilon > 0$ in the quadrant $Q_I$ and $\tau_\epsilon < 0$ in the
quadrant $Q_{II}$; this follows from the definition of $\tau_\epsilon$ and
from the positivity of $F_\epsilon$ (in fact, for $E_\epsilon \leq 0$, we have
$F_\epsilon \geq 1$). Hence, to the right (${\mathcal M}_+$) of the shell,
there is a part of $Q_{II}$ with the center to the right, and to the left of
the shell (${\mathcal M}_-$), there must be $Q_I$ with the center to the left,
Q. E. D.

An interpretation of the theorem is that the shell spacetime containing
one or two negative-mass subspacetimes is unphysical: there must be at least
one negative mass source somewhere outside the shell. The theorem also says
that a shell spacetime constructed from one flat and one positive-mass
subspacetimes must have a regular center inside the flat subspacetime.

To summarize: The action (\ref{act}) generates a regular dynamics in the
enlarged phase space defined just by
\be
  {\sf M} > 0,\quad R > 0.
\label{ran21}
\ee
For $E_+ > 0$ and $E_- >0$, it describes a shell interacting with a black hole
of mass $E_-$ (or $E_+$, depending on from where we observe the shell). For
$E_+ > 0$ and $E_- =0$ (or $E_+ = 0$ and $E_- >0$), it describes a
self-gravitating shell with flat spacetime inside. If any of $E_+$ and $E_-$
are negative, it describes the shell interacting with a negative mass
source. Only the points with
\be
  E_+ \geq 0,\quad E_- \geq 0
\label{ran22}
\ee 
are physically sensible; the subset specified by the inequalities
(\ref{ran21}) and (\ref{ran22}) is the {\em physical region} of the phase
space. There seems to be nothing about the variational principle (\ref{act})
that would help us to enforce the validity of the Eqs.\ (\ref{ran22}) in any
natural, automatic way.

\section{Radial and Kruskal momenta}
\label{sec:mom}
From the definition of the functions $T^\pm$ and $P$ as given in the previous
section, it follows that they are singular at the horizons: $T^\epsilon$ at
$R=2E_\epsilon$ and $P$ at $R=2E_-$ as well as at $R=2E_+$.  In this section,
we introduce three functions that are well-defined on the whole constraint
surface to replace $T^\pm$ and $P$. Another replacement of $T^\pm$ and $P$ with
similar properties has been tried in Ref.\ \cite{H1}. There, the functions were
constructed from the Kruskal coordinates; however, the Kruskal coordinates
make no sense for flat spacetime and so the important case $E_- = 0$ could not
be incorporated. In this subsection, we remove this problem.

Let us define the new functions $q$, $[T_1]$ and $\bar{T}_1$ by
\begin{eqnarray}
  q & := & a_-b_+b_-R\sqrt{|F_+F_-|}\text{sh}_{-a_+a_-}\frac{P}{R},
\label{defq} \\
  \bar{T}_1 & := & \frac{T^+ + T^-}{2} + E_+\ln\left|\frac{u_+}{v_+}\right| +
  E_-\ln\left|\frac{u_-}{v_-}\right|,
\label{defbarT1} \\
  \mbox{} [T_1] & := & T^+ - T^- +
  2E_+\ln\left|\frac{u_+}{v_+}\right| - 
  2E_-\ln\left|\frac{u_-}{v_-}\right|,
\label{def[T1]} 
\end{eqnarray}
and the auxiliary functions $u_\epsilon$ and $v_\epsilon$ by 
\be
  u_\epsilon := -\frac{q}{\sf M} -  \frac{[E]}{\sf M} + \epsilon\frac{\sf
  M}{2R},\quad v_\epsilon := -\frac{q}{\sf M} + \frac{[E]}{\sf M} -
  \epsilon\frac{\sf M}{2R},
\label{u} 
\ee
in Eq.\ (\ref{u}), the functions $q$ and $\sf M$ are given by Eqs.\ 
(\ref{defq}), (\ref{constr}) and $[E] := E_+ - E_-$.

The meaning of the momentum $q$ can be seen, if it is expressed by means of
velocities along dynamical trajectories. Comparing Eqs.\ (\ref{dotTm}),
(\ref{dotR}) and (\ref{defq}), we find that
\be
  q = {\sf M}\frac{dR}{d{\sf T}};
\label{meanq}
\ee
hence, $q$ is a kind of radial momentum, and it is regular everywhere along
dynamical trajectories. 

The meaning of the functions $\bar{T}_1$ and $[T_1]$ can be inferred from that
of $T_1^\epsilon$, which are defined by
\[
  T_1^\epsilon := \bar{T}_1 +\frac{\epsilon}{2}[T_1].
\]
Eqs.\ (\ref{defbarT1}) and (\ref{def[T1]}) lead then to
\[
  T_1^\epsilon = T^\epsilon +
  2E_\epsilon\ln\left|\frac{u_\epsilon}{v_\epsilon}\right| 
\]
for each $\epsilon = \pm 1$. Let us limit ourselves to the physical region and
distinguish two cases. \underline{$E_\epsilon = 0$}. Then, $T_1^\epsilon =
T^\epsilon$ in both quadrants of the $E$-spacetime, so $b_\epsilon
T_1^\epsilon$ is the time coordinate of the inertial system of the Minkowski
spacetime ${\mathcal M}_\epsilon$ in which the center of mass of the shell is
in rest. \underline{$E_\epsilon > 0$}. Then, it is possible to introduce
Kruskal coordinates $U_\epsilon$ and $V_\epsilon$ in the Kruskal spacetime
${\mathcal M}_\epsilon$ by
\bea
  R & = & 2E_\epsilon \kappa(-U_\epsilon V_\epsilon), \nn \\
  T^\epsilon & = & 2E_\epsilon\ln\left|\frac{V_\epsilon}{U_\epsilon}\right|
\label{TUV}
\eea
and by the requirement that both functions $U_\epsilon$ and $V_\epsilon$
increse to the future, where $\kappa$ is the well-known monotonous function
defined on the interval $(-1,\infty)$ by its inverse,
\[
  \kappa^{-1}(x) = (x-1)\text{e}^x.
\]
At each point of the shell trajectory $U_\epsilon = U_\epsilon(t)$ and
$V_\epsilon = V_\epsilon(t)$, the so-called {\em Kruskal momentum},
$P_K^\epsilon$, was defined in Ref.\ \cite{H1} by
\[
  P_K^\epsilon := \frac{R}{2}\ln\frac{dV_\epsilon}{dU_\epsilon}.
\]
In Ref.\ \cite{H2}, four further functions $u_\epsilon$ and $v_\epsilon$ were
defined by
\be
  u_\epsilon := \frac{U_\epsilon\text{e}^{P_K^\epsilon
    /R}}{\sqrt{\kappa_\epsilon\text{e}^{\kappa_\epsilon}}},\quad v_\epsilon :=
    \frac{V_\epsilon\text{e}^{-P_K^\epsilon 
    /R}}{\sqrt{\kappa_\epsilon\text{e}^{\kappa_\epsilon}}},
\label{defuv2}
\ee
where $\kappa_\epsilon := \kappa(-U_\epsilon V_\epsilon)$.

The following relation between the pairs $u_\epsilon, v_\epsilon$ and $R,
E_\epsilon$ was shown to hold in Ref.\ \cite{H2} (Eqs.\ (32) and (33)): 
\beann
  u_\epsilon & = & -\frac{[E]}{\sf M} + \epsilon\frac{\sf M}{2R}
  +\omega\sqrt{-V}, \\
  v_\epsilon & = & \frac{[E]}{\sf M} - \epsilon\frac{\sf M}{2R}
  +\omega\sqrt{-V}.  
\eeann 
A comparison with Eqs.\ (\ref{radeq}) and (\ref{radpot}) shows that our
functions $u_\epsilon$ and $v_\epsilon$ defined by Eqs.\ (\ref{u}) coincide
with the functions $u_\epsilon$ and $v_\epsilon$ of Ref.\ \cite{H2}, if the
dynamical equations are satisfied. Eq.\ (\ref{defuv2}) then shows that
$u_\epsilon$ and $v_\epsilon$ vanish at horizons and that their logarithms
diverge there. It then also follows for the functions $T_1^\epsilon$ along
dynamical trajectories that
\be 
  T_1^\epsilon = \frac{4E_\epsilon P_K^\epsilon}{R}.
\label{meanT1}
\ee
We can see that the functions $T_1^\epsilon$ are directly determined by the
geometry of the dynamical trajectory in the spacetime ${\mathcal M}_\epsilon$
and are regular everywhere along the trajectory.

\section{A regular coordinate system}
\label{sec:coord}
In this section, we describe a coordinate system that is regular at the
horizons, and at the same time remains meaningful for the special values
$E_\epsilon = 0$ of Schwarzschild masses.

As such coordinates on $\Gamma$, we suggest the seven functions $q$, $R$, $\sf
M$, $\sf T$, $[E]$, $\bar{T}_1$, and $[T_1]$. Let us show that the Jacobian of
the transformation from these coordinates to $P$, $R$, $E_+$, $T^+$, $E_-$,
$T^-$ and $\sf T$ is non-zero. To calculate the determinant, we observe that
the four columns $R$, $T^+$, $T^-$ and $\sf T$ of the Jacobian
\[
  \frac{\partial(q,\ldots,[T_1])}{\partial(P,\ldots,{\sf T})}
\]
contain each at most two non-zero elements; if we expand the determinant
along these columns, we obtain
\be
  \frac{\partial(q,\ldots,[T_1])}{\partial(P,\ldots,{\sf T})} =
  -\frac{\partial q}{\partial P}\left(\frac{\partial {\sf M}}{\partial
      E_+} + 
    \frac{\partial {\sf M}}{\partial E_-}\right) + \frac{\partial {\sf
      M}}{\partial P}\left(\frac{\partial q}{\partial E_+} + \frac{\partial
      q}{\partial E_-}\right). 
\label{jacob}
\ee
For the derivatives involved here, Eqs.\ (\ref{constr}) and (\ref{defq}) yield
immediately: 
\[
  \frac{\partial q}{\partial P} = 
  a_-b_+b_-\sqrt{|F_+F_-|}\text{sh}_{a_+a_-}\frac{P}{R},\quad 
  \frac{\partial {\sf M}}{\partial P} =
  -a_-b_+b_-\sqrt{|F_+F_-|}\frac{R}{\sf M}\text{sh}_{-a_+a_-}\frac{P}{R}, 
\]
\begin{eqnarray*} 
  \frac{\partial {\sf M}}{\partial E_+} + \frac{\partial {\sf M}}{\partial
  E_-} & = & \frac{R}{\sf M}\left(-2 + a_+b_+b_-\frac{F_+ +
  F_-}{\sqrt{|F_+F_-|}}\text{sh}_{a_+a_-}\frac{P}{R}\right), \\
  \frac{\partial q}{\partial E_+} + \frac{\partial q}{\partial E_-} & = & 
  -a_+b_+b_-\frac{F_+ +
  F_-}{\sqrt{|F_+F_-|}}\text{sh}_{-a_+a_-}\frac{P}{R}.
\end{eqnarray*}
Substituting this in Eq.\ (\ref{jacob}) and using again Eq.\ (\ref{constr}), we
arrive at the simple result
\[
  \frac{\partial(q,\ldots,[T_1])}{\partial(P,\ldots,{\sf T})} = -\frac{\sf
  M}{R}, 
\]
which holds in each sector of $\Gamma$.

It follows that the transformation is regular within each sector and can be
inverted. In fact, it is not difficult to find the functions defining the
inverse transformation. First, we show that
\be 
  \bar{E} = \frac{R}{2}\left(1 + \frac{q^2}{{\sf M}^2} - \frac{[E]^2}{{\sf
      M}^2} -\frac{{\sf M}^2}{4R^2}\right)
\label{barM}
\ee
in each sector. To this aim, we substitute for $q$ and $\sf M$ from Eqs.\
(\ref{constr}) and (\ref{defq}) into Eq.\ (\ref{barM}) and use the identities
\be
  F_+ + F_- = 2\left(1 - \frac{2\bar{E}}{R}\right), \quad
  F_+F_- = \left(1 - \frac{2\bar{E}}{R}\right)^2 - \frac{[E]^2}{R^2}.
\label{FF}
\ee
In this way, we obtain
\[
  \frac{2\bar{E}}{R} - 1 = \frac{R^2}{{\sf M}^2}\left(-2(1-2\bar{E}/R)^2 +
  2a_-b_+b_-(1-2\bar{E}/R)\sqrt{|F_+F_-|}\text{sh}_{a_+a_-}\frac{P}{R}\right). 
\]
Now, the equality follows immediately from Eq.\ (\ref{constr}).

The next non-trivial part of the inverse transformation is the relation
\be
  P = \frac{R}{2}\ln\left|\frac{u_+v_-}{u_-v_+}\right|,
\label{P}
\ee
which again holds in each sector. To show Eq.\ (\ref{P}), we use Eqs.\
(\ref{u}) to obtain 
\[
  \frac{u_+v_-}{u_-v_+} = \frac{\left(q - \frac{{\sf M}^2}{2R}\right)^2 -
  [E]^2}{\left(q + \frac{{\sf M}^2}{2R}\right)^2 - [E]^2}.
\]
The substitution for $q$ and $\sf M$ gives
\[
  q + \eta \frac{{\sf M}^2}{2R} = \eta(R - 2\bar{E}) - \eta
  a_{-\eta}b_+b_-R\sqrt{|F_+F_-|}\text{e}^{\eta P/R},
\]
where $\eta = \pm 1$. After some rearrangement and applying Eqs.\ (\ref{FF}),
we have 
\[
  \left(q + \eta \frac{{\sf M}^2}{2R}\right)^2 - [E]^2 =
  -a_{-\eta}b_+b_-R^2\sqrt{|F_+F_-|}\text{e}^{\eta P/R} \left(F_+ + F_- -
  2a_-b_+b_+\text{sh}_{a_+a_-}\frac{P}{R}\right). 
\]
Then, Eq.\ (\ref{constr}) leads immediately to
\[
  \left(q + \eta \frac{{\sf M}^2}{2R}\right)^2 - [E]^2 = -a_{\eta}b_+b_-{\sf
  M}^2\sqrt{|F_+F_-|}\text{e}^{\eta P/R},
\]
and this implies Eq.\ (\ref{P}).

The rest of the inverse transformation is easy: the functions $R$ and $\sf T$
are the same in both sets of variables,
\be
  E_\epsilon = \bar{E} + \frac{\epsilon}{2}[E]
\label{Meps}
\ee
and 
\be
  T^\epsilon = \bar{T}_1 + \frac{\epsilon}{2}[T_1] -
  2E_\epsilon\ln\left|\frac{u_\epsilon}{v_\epsilon}\right|;
\label{Teps}
\ee
in the last equation, one has to substitute Eq.\ (\ref{Meps}) for $E_\epsilon$
and Eqs.\ (\ref{u}) for $u_\epsilon$ and $v_\epsilon$. 

The ranges of the variables $q$, $R$, $\sf M$, $\sf T$, $[E]$, $\bar{T}_1$,
and $[T_1]$ in the physical region are implied by the conditions (\ref{ran21})
(\ref{ran22}) and Eq.\ (\ref{barM}). Let us work them out. Eq.\ (\ref{barM})
together with $\bar{E} \geq 0$ implies that
\be
  [E]^2 \leq {\sf M}^2 + q^2 - \frac{{\sf M}^4}{4R^2};
\label{ran[M]}
\ee
hence,
\be
  {\sf M}^2 + q^2 \geq \frac{{\sf M}^4}{4R^2}.
\label{ranq}
\ee
The two inequalities $E_+ \geq 0$ and $E_- \geq 0$ are equivalent to $[E]^2
\leq 4\bar{E}^2$. Eq.\ (\ref{barM}) implies that this inequality is, in turn,
equivalent to
\be
  \Biggl([E]^2 - \biggl({\sf M}^2 + q^2 + \frac{{\sf
      M}^4}{4R^2}\biggr)\Biggr)^2 \geq \left(\frac{{\sf M}^2}{R}\sqrt{{\sf
        M}^2 + q^2}\right)^2.
\label{5.1}
\ee
From Eq.\ (\ref{ran[M]}), it follows that
\[
  \mbox{} [E]^2 - \biggl({\sf M}^2 + q^2 + \frac{{\sf
      M}^4}{4R^2}\biggr) \leq - \frac{{\sf M}^4}{2R^2} < 0.
\]
Eq.\ (\ref{5.1}) is, therefore, equivalent to
\be
  [E]^2 \leq \left(\sqrt{{\sf M}^2 + q^2} - \frac{{\sf M}^2}{2R}\right)^2.
\label{5.2}
\ee
Eq.\ (\ref{ranq}) shows that the inequality (\ref{5.2}) is not weaker than
(\ref{ran[M]}). Hence, all information is contained in the following
inequality: 
\be
  |[E]| \leq \sqrt{{\sf M}^2 + q^2} - \frac{{\sf M}^2}{2R}.
\label{ranM}
\ee
Inequality (\ref{ranM}) together with ${\sf M} > 0$ and $R > 0$ define
the ranges of the variables $q$, $R$, $\sf M$, $\sf T$, $[E]$, $\bar{T}_1$,
and $[T_1]$ in the physical region.

\section{Symmetry of the system}
\label{sec:symm}
In this section, we collect the results of Ref.\ \cite{H2} on the symmetry of
the shell action and describe the action of the symmetry transformations on
the new variables.

In Ref.\ \cite{H2}, we have found that there is a continuous symmetry group
generated by an arbitrary function of $E_+$ and $E_-$. A simple generalization
of the argument given in Ref.\ \cite{H2} implies that the following finite
transformation of the variables $E_\epsilon$, $u_\epsilon$, $v_\epsilon$ and
$\tilde{P}_\epsilon$, which were used there, is a symmetry of the system:
\[
  \tilde{P}_\epsilon \mapsto \tilde{P}_\epsilon
  -\epsilon\frac{\partial\Lambda}{4\partial E_\epsilon},\quad \epsilon = \pm
  1,
\]
the other variables being invariant, where $\Lambda(E_+,E_-)$ is an arbitrary
smooth function of two variables (the factor 1/4 is introduced for
convenience). In addition to this continuous infinitely dimensional group,
there were two reflections, which we denote here by $\sigma_1$ and $\sigma_2$.
$\sigma_1$ is a time reflection defined by
\[
  E_\epsilon \mapsto E_\epsilon,\quad u_\epsilon + u_\epsilon \mapsto
  -(u_\epsilon + v_\epsilon),\quad u_\epsilon - u_\epsilon \mapsto
  u_\epsilon - v_\epsilon,\quad \tilde{P}_\epsilon \mapsto
  -\tilde{P}_\epsilon,
\]
and $\sigma_2$ is a left-right reflection defined by
\[
  E_+ \leftrightarrow E_-,\quad u_+ \leftrightarrow v_-,\quad u_-
  \leftrightarrow v_+,\quad \tilde{P}_+ \leftrightarrow 
  -\tilde{P}_-.
\]

These results can be easily expressed in our variables $q$, $R$, $[E]$,
$\bar{T}_1$, $\bar{E}$ and $[T_1]$ and extended to $\sf M$ and $\sf T$.
Consider the function $q$; Eqs.\ (\ref{u}) yield
\[
  q = -{\sf M}(u_\epsilon + v_\epsilon),\quad \epsilon = \pm 1.
\]
It is clear that $\sf M$ is invariant with respect of the whole group; hence,
$q$ is invariant with respect to the $\Lambda$-transformation for any
$\Lambda(E_+,E_-)$, it changes sign by the time reflection $\sigma_1$ and is
invariant with respect to $\sigma_2$. The function $R$ was written in terms of
$u_\epsilon$ and $v_\epsilon$ in Ref.\ \cite{H2} in the form
\[
  R = \frac{2E_\epsilon}{1+u_\epsilon v_\epsilon},\quad \epsilon = \pm 1.
\]
Realising that $4u_\epsilon v_\epsilon = (u_\epsilon + v_\epsilon)^2 -
(u_\epsilon - v_\epsilon)^2$, we can see that $R$ is an invariant of all above
transformations. The function $[E]$ transforms non-trivially only by
$\sigma_2$,
\[
  [E] \mapsto -[E],
\]
and $\bar{E}$ is an invariant like $R$. A comparison of our Eq.\
(\ref{meanT1}) with Eq.\ (43) of Ref.\ \cite{H2} reveals that $T_1^\epsilon =
4\tilde{P}_\epsilon$, so that its $\Lambda$-transformation is
\[
  T_1^\epsilon \mapsto T_1^\epsilon -\epsilon\frac{\partial \Lambda}{\partial
  E_\epsilon}. 
\]
Using this formula, one finds easily that 
\be
  \bar{T}_1 \mapsto \bar{T}_1 - \frac{\partial \Lambda}{\partial [E]},\quad
  [T_1] \mapsto [T_1] - \frac{\partial \Lambda}{\partial \bar{E}}.
\label{lambdaT}
\ee
The time reflection $\sigma_1$ changes the signs of both times:
\[
  \bar{T}_1 \mapsto -\bar{T}_1,\quad [T_1] \mapsto -[T_1],
\]
but the left-right reflection $\sigma_2$ acts as follows:
\[
  \bar{T}_1 \mapsto -\bar{T}_1,\quad [T_1] \mapsto [T_1].
\]
From the physical meaning of
$\sf M$ and $\sf T$, it follows that they are invariant with respect to
the continuous transformation for any $\Lambda(E_+,E_-)$ and of the left-right
reflection $\sigma_2$, whereas the time reflection $\sigma_1$ must give
\[
  {\sf M} \mapsto {\sf M},\quad {\sf T} \mapsto -{\sf T}.
\]
However, as $\sf T$ is a new cyclic coordinate, the new system has larger
continuous symmetry: we can extend $\Lambda$ to an arbitrary function of
the three variables $E_+$, $E_-$, and $\sf M$, and define the action on $\sf
T$ as follows:
\be
  {\sf T} \mapsto {\sf T} + \frac{\partial\Lambda}{\partial{\sf M}}.
\label{lambdasfT}
\ee
Let us postpone the proof that this extended $\Lambda$-transformation is a
symmetry to Sec.\ \ref{sec:cartan}. It generates three independent constant
shifts of the cyclic variables $\bar{T}_1$, $[T_1]$ and $\sf T$ that are
different in different shell spacetimes.

We can see from these results that our new variables transform particularly
simply. This was, in fact, the idea that helped to find the variables in the
first place.

\section{Transformations of the Cartan form}
\label{sec:cartan}
Let us transform the Cartan form $\Theta$ in each sector of $\Gamma$ to the
variables $q$, $R$, $\sf M$, $\sf T$, $[E]$, $\bar{T}_1$ and $[T_1]$ and check
that it can then be extended smoothly accross the boundaries of the sectors.

If we substitute for $P$, $T^+$, $T^-$ and $\sf M$, we obtain
\begin{eqnarray*}
  \Theta & = & -\bar{E}d[T_1] - [E]d\bar{T}_1 + {\sf M}d{\sf T} + 2E_+^2
  d\left(\ln\left|\frac{u_+}{v_+}\right|\right) - 2E_-^2 
  d\left(\ln\left|\frac{u_-}{v_-}\right|\right) \\
  & & \mbox{} + \ln\left|\frac{u_+v_-}{u_-v_+}\right|
  d\left(\frac{R^2}{4}\right) + 
  \ln \left|\frac{u_+}{v_+}\right| d (E_+^2) - \ln
  \left|\frac{u_-}{v_-}\right| d (E_-^2). 
\end{eqnarray*}
The logarithms can be rearanged as follows:
\begin{eqnarray*}
  \Theta & = & -\bar{E}d[T_1] - [E]d\bar{T}_1 + {\sf M}d{\sf T} +
  d\Biggl(\biggl(\frac{R^2}{4} + E_+^2\biggr)\ln\left|\frac{u_+}{v_+}\right| - 
  \biggl(\frac{R^2}{4} + E_-^2\biggr)\ln\left|\frac{u_-}{v_-}\right|\Biggr) \\
  & & \mbox{} - \left(\frac{R^2}{4} -
  E_+^2\right)d\left(\ln\left|\frac{u_+}{v_+}\right|\right) +
  \left(\frac{R^2}{4} - 
  E_-^2\right)d\left(\ln\left|\frac{u_-}{v_-}\right|\right). 
\end{eqnarray*}
The total differential of a function (that diverges badly at the horizons) can
be left out. We obtain
\be
  \Theta = -\bar{E}d[T_1] - [E]d\bar{T}_1 + {\sf M}d{\sf T} 
  - \left[\left(\frac{R^2}{4} -
  E^2\right)d\left(\ln\left|\frac{u}{v}\right|\right)\right],
\label{theta2}
\ee
where we must substitute for $E_+$, $E_-$, $\bar{E}$, $u_\epsilon$ and
$v_\epsilon$ from Eqs.\ (\ref{Meps}), (\ref{barM}) and (\ref{u}). This form of
$\Theta$ has been obtained in Ref.\ \cite{H2} (without the matter term); there,
however, $u_\epsilon$ and $v_\epsilon$ were regarded as independent
variables. Observe that Eqs.\ (\ref{u}), (\ref{barM}) and (\ref{Meps}) imply 
\be
  \frac{2E_\epsilon}{R} - 1 = u_\epsilon v_\epsilon
\label{uv}
\ee
so that $u_\epsilon$ and $v_\epsilon$ are not independent functions for
$E_\epsilon = 0$. 

Now, we can show that the last term in $\Theta$ is regular: we just rewrite
it using Eq.\ (\ref{uv}):
\be
  \left[\left(\frac{R^2}{4} -
  E^2\right)d\left(\ln\left|\frac{u}{v}\right|\right)\right] = 
  \left[\frac{R^2}{4}\left(1 + \frac{2E}{R}\right)(vdu - udv)\right].
\label{log}
\ee 
Thus, our new variables cover all of the constraint hypersurface
$\Gamma$ as promised.

Let us express $\Theta$ explicitly by means of the variables $q$, $R$, $\sf M$
and $[E]$. Employing Eqs.\ (\ref{u}), we have
\begin{eqnarray*}
  \overline{vdu - udv} & = & 2\left(-\frac{[E]}{{\sf M}^2}dq + \frac{q}{{\sf
  M}^2}d[E]\right), \\
  \mbox{} [vdu - udv] & = & 2\left(\frac{1}{R}dq + \frac{q}{R^2}dR -
  \frac{2q}{R{\sf M}}d{\sf M}\right).
\end{eqnarray*}
Then, applying the well-known identity $[AB] = \bar{A}[B] + \bar{B}[A]$
that holds for jumps and averages of any two functions $A$ and $B$, and after
using Eq.\ (\ref{barM}), we arrive at
\begin{eqnarray}
  \lefteqn{\left[\left(\frac{R^2}{4} -
  E^2\right)d\left(\ln\left|\frac{u}{v}\right|\right)\right] =
  \frac{Rq[E]}{{\sf M}^2}d[E] - \left(\frac{2Rq}{\sf M} + 
  \frac{Rq^3}{{\sf M}^3} -\frac{Rq[E]^2}{{\sf M}^3} - \frac{q{\sf
  M}}{4R}\right)d{\sf M} } \nn \\ 
  & & \mbox{} + \left(R
  +\frac{Rq^2}{2{\sf M}^2} - \frac{3R[E]^2}{2{\sf M}^2} - \frac{{\sf
  M}^2}{8R}\right)dq + \left(q + \frac{q^3}{2{\sf M}^2} -
  \frac{q[E]^2}{2{\sf M}^2} - \frac{q{\sf M}^2}{8R^2}\right)dR.
\label{resthet}
\end{eqnarray}
This, together with Eq.\ (\ref{theta2}) gives $\Theta$ as a function of new
variables. It is a complicated one, so we have to transform the coordinates
further in order to get a simple expression.

The following question will give some direction to this search for simplicity:
is the variable $[T_1]$ suitable for the role of time? More concretely, the
time levels in the constraint surface should be {\em transversal} (this is
shown in Ref.\ \cite{H3}). A surface is transversal, if dynamical 
trajectories intersect it transversally and only once. We can see if the
surface $[T_1]=$ const has this property as follows. The direction of motion
on the constraint surface coincides with the direction of degeneration of the
presymplectic form $d\Theta$; hence, the pull-back $d\Theta_f$ of
$d\Theta$ to any transversal surface $f =$ const must be non-degenerate. If
the surface is $2n$-dimensional, then the $2n$-form
$d\Theta_f\wedge\ldots\wedge d\Theta_f$ must be everywhere non-zero. In our
case, $n = 3$. Let us calculate $d\Theta_{[T_1]}\wedge d\Theta_{[T_1]}\wedge
d\Theta_{[T_1]}$. Eqs.\ (\ref{theta2}) and (\ref{resthet}) imply that
\beann
  \lefteqn{d\Theta_{[T_1]} = -d[E]\wedge d\bar{T}_1 + d{\sf M}\wedge d{\sf T}
    } \\
  && \mbox{} + A_1dR\wedge dq + A_2 dR\wedge d[E] + A_3 dR\wedge d{\sf M} +
  A_4dq\wedge d[E] + A_5dq\wedge d{\sf M} + A_6d[E]\wedge d{\sf M},
\eeann
where $A_i$, $i = 1,\ldots,6$ are some functions of the variables $R$, $q$,
$[E]$, and $\sf M$. It is clear, therefore, that
\[
  d\Theta_{[T_1]}\wedge d\Theta_{[T_1]}\wedge d\Theta_{[T_1]} =
  -6A_1d[E]\wedge 
  d\bar{T}_1\wedge d{\sf M}\wedge d{\sf T}\wedge dR\wedge dq,
\]
and the points where the six-form vanishes coincide with those where $A_1$
vanishes. $A_1$ can be calculated from Eq.\ (\ref{resthet}) with the result
\be
  A_1 = \frac{q^2}{{\sf M}^2} + \frac{[E]^2}{{\sf M}^2} - \frac{{\sf
      M}^2}{4R^2}. 
\label{A1}
\ee
Thus, $A_1$
vanishes for 
\[ \mbox{}
  [E]^2 = \frac{{\sf M}^4}{4R^2} - q^2.
\]
Do these points lie in the physical region that is given by inequality
(\ref{ranM})? Let us substitute the value of $[E]^2$ into Eq.\ (\ref{ranM})
and rewrite the result in the form
\[
  \frac{{\sf M}^2}{R^2} \leq 4\left(1 + \frac{q^2}{{\sf M}^2}\right) -
  \frac{3{\sf M}^2}{{\sf M}^2 + q^2}.
\]
This inequality implies inequality (\ref{ranq}); hence, all zero points of
$A_1$ lie in the physical region. We conclude that $[T_1] =$ const is {\em no} 
transversal surface.

Let us try to shift $[T_1]$ by a function dependent on the variables $R$, $q$,
$[E]$ and $\sf M$:
\[
  [T_1] = T_2 + X(R,q,[E],{\sf M}).
\]
Our motivation for choosing such a form is that $T_2$ is then transformed
by the $\Lambda$-symmetry (\ref{lambdaT}) in the same way as $[T_1]$ is, and
that it is also a cyclic coordinate. Substitution for $[T_1]$ into
$d\Theta$ changes the coefficient $A_1$ at $dR\wedge dq$ in $d\Theta_{T_2}$ by
\[
  A_1 \rightarrow A_1 + \frac{\partial\bar{E}}{\partial q}\frac{\partial
  X}{\partial R} - \frac{\partial\bar{E}}{\partial R}\frac{\partial
  X}{\partial q}.
\]
Comparing $A_1$, Eq.\ (\ref{A1}), with $\partial\bar{E}/\partial R$ calculated
from Eq.\ (\ref{barM}),
\[
  \frac{\partial\bar{E}}{\partial R} = \frac{1}{2} + \frac{q^2}{2{\sf M}^2} -
  \frac{[E]^2}{2{\sf M}^2} + \frac{{\sf M}^2}{8R^2},
\]
we observe that
\[
  A_1 + 2\frac{\partial\bar{E}}{\partial R} = 2 + 2\frac{q^2}{{\sf M}^2},
\]
which is always positive; thus, one possible shift is
\be
  [T_1] = T_2 - 2q.
\label{shiftT2}
\ee
If we perform this transformation in $\Theta$, we find out easily that another
shift,
\be
  \bar{T}_1 = T_3 + 2\frac{Rq[E]}{{\sf M}^2},
\label{shiftT3}
\ee
miraculously cancels some pesky cross terms in $d\Theta$. Indeed, shifting the
`times' in the Cartan form, we have: 
\begin{eqnarray*}
  \lefteqn{-\bar{E}d[T_1] - [E]d\bar{T}_1 + {\sf M}d{\sf T} = -\bar{E}dT_2 -
  [E]dT_3 + {\sf M}d{\sf T}} \\
  & & \mbox{} - \left(R + \frac{Rq^2}{{\sf M}^2} - \frac{3R[E]^2}{{\sf M}^2} -
  \frac{{\sf M}^2}{4R}\right)dq + \frac{2q[E]^2}{{\sf M}^2}dR +
  \frac{2qR[E]}{{\sf M}^2}d[E] - \frac{4qR[E]^2}{{\sf M}^3}d{\sf M}.
\end{eqnarray*}
Using Eqs.\ (\ref{theta2}) and (\ref{resthet}), we obtain
\begin{eqnarray*}
  \Theta & = & -\bar{E}dT_2 - [E]dT_3 + {\sf M}d{\sf T} -
  d\left(\frac{3qR[E]^2}{2{\sf M}^2} + \frac{q{\sf M}^2}{8R}\right) \\
  & & \mbox{} + \frac{Rq^2}{2{\sf M}^2}dq - \left(q + \frac{q^3}{2{\sf
  M}^2}\right)dR + \frac{2R}{\sf M}\left(q + \frac{q^3}{2{\sf
  M}^2}\right)d{\sf M}.
\end{eqnarray*}
The last three terms do not contain other variables than $q$, $R$ and ${\sf
  M}$ and so the terms with $d[E]\wedge dR$, $d[E]\wedge dq$ and $d[E]\wedge
d{\sf M}$ disappear from $d\Theta$. We have not found any simple physical or
geometrical interpretation of the new times $T_2$ and $T_3$; observe that the
shifts (\ref{shiftT2}) and (\ref{shiftT3}) ``mix the sides''.

The fact that the last three terms in $\Theta$ contain only three variables
means that they can be simplified to just one term. For example, 
\[
  \frac{Rq^2}{2{\sf M}^2}dq - \left(q + \frac{q^3}{2{\sf
  M}^2}\right)dR + \frac{2R}{\sf M}\left(q + \frac{q^3}{2{\sf
  M}^2}\right)d{\sf M} = 
  d\left(\frac{Rq^3}{6{\sf M}^2}\right) - \left(q{\sf M}^2 +
  \frac{2q^3}{3}\right)d\left(\frac{R}{{\sf M}^2}\right). 
\]
This suggests the second step of our transformation:
\begin{eqnarray}
  p & := & q{\sf M}^2 + \frac{2q^3}{3},
\label{p} \\
  x& := & \frac{R}{{\sf M}^2}.
\label{x}
\end{eqnarray}
We arrive so at the final shape of the Cartan form
\be
  \Theta = -\bar{E}dT_2 - [E]dT_3 + {\sf M}d{\sf T} - pdx,
\label{newtheta}
\ee
where $\bar{E}$ is given by Eq.\ (\ref{barM}) in which $q$ and $R$ are
expressed in terms of $p$ and $x$; such expressions can be obtained  by
solving Eqs.\ (\ref{p}) and (\ref{x}) for $q$ and $R$. 

Consider Eq.\ (\ref{p}). If we keep $\sf M$ fixed, $p$ is an increasing
function of $q$ in the whole interval $(-\infty,\infty)$, for
\[
  \frac{\partial p}{\partial q} = {\sf M}^2 + 2q^2 > 0.
\]
Hence, the function maps the range $(-\infty,\infty)$ of $q$ onto the range
$(-\infty,\infty)$ of $p$ in a bijective way, and it possesses a unique
inverse. One can explicitly write down this inverse using second and third
roots; this will make $\bar{E}$ a complicated function of $p$, $x$, $\sf M$
and $[E]$ (we shall write down this function later, cf.\ Eq.\ (\ref{solEG7})).

Eq.\ (\ref{newtheta}) implies that
\beann
  d\Theta_{T_2}\wedge d\Theta_{T_2}\wedge d\Theta_{T_2} & = & 6d[E]\wedge
  dT_3\wedge d{\sf M}\wedge d{\sf T}\wedge dp\wedge dx, \\
  d\Theta_{T_3}\wedge d\Theta_{T_3}\wedge d\Theta_{T_3} & = & 6\frac{\partial
    \bar{E}}{\partial [E]} d[E]\wedge
  dT_2\wedge d{\sf M}\wedge d{\sf T}\wedge dp\wedge dx, \\
  d\Theta_{T}\wedge d\Theta_{T}\wedge d\Theta_{T} & = & - 6\frac{\partial
    \bar{E}}{\partial {\sf M}}d[E]\wedge
  dT_3\wedge d{\sf M}\wedge d{T_2}\wedge dp\wedge dx, 
\eeann
and Eq.\ (\ref{barM}) gives
\beann
  \frac{\partial \bar{E}}{\partial [E]} & = & -\frac{R}{{\sf M}^2}[E], \\
  \frac{\partial \bar{E}}{\partial {\sf M}} & = & -\frac{\sf M}{4R}.
\eeann
Thus, the surface $T_3 =$ const is not transversal at $[E] = 0$, but $T_2=$
const and ${\sf T} =$ const are transversal everywhere.

The Cartan form (\ref{newtheta}) is not yet very simple, because $\bar{E}$ is
a complicated function of $p$, $x$, $[E]$ and $\sf M$. What we can still do is
to extend the phase space to an eight-dimensional manifold with the cordinates
$\bar{E}$, $[T_2]$, $[E]$, $\bar{T}_2$, $\sf M$, $\sf T$, $p$ and $x$, and to
express the constraint between these variables as a polynomial. Indeed,
applying Eq.\ (\ref{x}), we can rewrite Eq.\ (\ref{barM}) as follows
\[
  q^2 = \frac{1}{4x^2} + \frac{2\bar{E}}{x} + [E]^2 - {\sf M}^2;
\]
squaring Eq.\ (\ref{p}) and substituting the above expression
for $q^2$ into the result, we obtain
\be
  p^2 = \frac{1}{9}\left(\frac{1}{4x^2} + \frac{2\bar{E}}{x} + [E]^2 - {\sf
      M}^2\right) \left(\frac{1}{2x^2} + \frac{4\bar{E}}{x} + 2[E]^2 + {\sf
      M}^2\right)^2.
\label{newconstr}
\ee 
This is a constraint that {\em is} polynomial in the momenta $p$,
$\bar{E}$, $[E]$ and $\sf M$. The inequalities defining the ranges of the
variables $\bar{E}$, $[T_2]$, $[E]$, $\bar{T}_2$, $\sf M$, $\sf T$, $p$ and
$x$ in the physical region are also simplified: 
\be
  [E]^2 \leq 4\bar{E}^2,\quad {\sf M} > 0,\quad x > 0.
\label{ran3}
\ee

Let us, finally, find the transformation of the variables $p$, $x$, $\sf M$,
$\sf T$, $\bar{E}$, $T_2$, $[E]$ and $T_3$ by the symmetry
group. The results are as follows. We have for the $\Lambda$-transformation,
\[
  p \mapsto p,\quad x \mapsto x,\quad {\sf M} \mapsto {\sf M},\quad
  {\sf T} \mapsto {\sf T},\quad \bar{E} \mapsto \bar{E},\quad 
  [E] \mapsto [E],
\]
\[
  T_2 \mapsto T_2 - \frac{\partial \Lambda}{\partial\bar{E}},\quad
  T_3 \mapsto T_3 - \frac{\partial \Lambda}{\partial[E]},\quad
  {\sf T} \mapsto {\sf T} + \frac{\partial \Lambda}{\partial{\sf M}};
\]
for the time reflection $\sigma_1$,
\[
  p \mapsto -p,\quad x \mapsto x,\quad {\sf M} \mapsto {\sf M},\quad
  {\sf T} \mapsto -{\sf T},\quad \bar{E} \mapsto \bar{E},\quad 
  [E] \mapsto [E],\quad T_2 \mapsto -T_2,\quad T_3 \mapsto -T_3;
\]
and for the left-right reflection $\sigma_2$,
\[
  p \mapsto p,\quad x \mapsto x,\quad {\sf M} \mapsto {\sf M},\quad
  {\sf T} \mapsto {\sf T},\quad \bar{E} \mapsto \bar{E},\quad 
  [E] \mapsto -[E],\quad T_2 \mapsto T_2,\quad T_3 \mapsto -T_3.
\]
Thus, the simple form of the symmetry transformations is preserved by the
shifts (\ref{shiftT2}) and (\ref{shiftT3}) as well as by transformations
(\ref{p}) and (\ref{x}). Eq.\ (\ref{newtheta}) implies that
\[
  \Theta \mapsto \Theta +
  d\left(\frac{\partial\Lambda}{\partial\bar{E}}\bar{E} +
  \frac{\partial\Lambda}{\partial [E]}[E] +
  \frac{\partial\Lambda}{\partial{\sf M}}{\sf E} - \Lambda\right)
\]
by a $\Lambda$-transformation, so our extended $\Lambda$-transformation {\em
  is} a symmetry,
\[
  \Theta \mapsto -\Theta
\]
by the time reflection $\sigma_1$,\footnote{The sign change of $\Theta$ by
  $\sigma_1$ is necessary, because time reflections are anti-symplectic
  trans\-form\-a\-tions.} and 
\[
  \Theta \mapsto \Theta
\]
by the left-right reflection $\sigma_2$. The constraint equation
(\ref{newconstr}) is clearly invariant of the group. Thus, the action of the
symmetry group is well-defined in the whole phase space and it preserves the
physical region.

\section{The constraint surface}
\label{sec:constr}
In this section, we are going to study the properties of the constraint
(\ref{newconstr}). We are interested in the topology of the constraint
surface; we shall investigate the salient question of whether or not any
spurious solutions have been unintentionally included during the process of
transforming the constraint to the polynomial form; and we shall look for
solutions of the constraint (\ref{newconstr}) with respect to some momenta.

Our new variational principle reads
\be
  S = \int dt(-\bar{E}\dot{T}_2 -[E]\dot{T}_3 +{\sf M}\dot{\sf T} - p\dot{x} -
  \nu_1{\mathcal C}_1),
\label{newaction}
\ee
where
\be
  {\mathcal C}_1 = 9p^2 - \left(\frac{1}{4x^2} + \frac{2\bar{E}}{x} +[E]^2 -
    {\sf M}^2\right)\left(\frac{1}{2x^2} + \frac{4\bar{E}}{x} + 2[E]^2 +
    {\sf M}^2\right)^2.
\label{newcf}
\ee
The constraint surface is trivial in the direction of the three times $T_2$,
$T_3$ and $\sf T$, so we can limit ourselves to its projection to the
five-dimensional space spanned by $\bar{E}$, $[E]$, $\sf M$, $p$ and $x$. Let
us observe that ${\mathcal C}_1$ depends on $\bar{E}$, $[E]$ and $x$ through
a function that we shall call $B$,
\be
  B := \frac{1}{4x^2} + \frac{2\bar{E}}{x} +[E]^2,
\label{B}
\ee
and that the gradient of $B$ is non-zero. It is, therefore, advantageous, to
write ${\mathcal C}_1$ as a composed function,
\[
 {\mathcal C}_1 = 9p^2 - (B - {\sf M}^2)(2B + {\sf M}^2)^2.
\]

Let us first look for the singular points of the constraint surface (that is,
where the gradient of ${\mathcal C}_1$ vanishes). We have
\bea
  \frac{\partial {\mathcal C}_1}{\partial p} & = & 18 p,
\label{dCdp} \\
  \frac{\partial {\mathcal C}_1}{\partial {\sf M}} & = & 6 {\sf M}^3(2B + {\sf
    M}^2), 
\label{dCdM} \\
  \frac{\partial {\mathcal C}_1}{\partial B} & = & -3(2B + {\sf M}^2)(2B -
  {\sf M}^2).
\label{dCdB}
\eea
As all three derivatives must vanish at a singular point, all such points are
determined by the equations
\be
  p = 0,\quad 2B + {\sf M}^2 = 0.
\label{static}
\ee
All points satisfying Eqs.\ (\ref{static}) are solutions of Eq.\
(\ref{newconstr}); however, the physical ranges (\ref{ran3}) of the variables
$\bar{E}$, $[E]$ and $x$ do not allow $B$ to be negative. We conclude that the
constraint surface is regular (smooth) in the physical region (\ref{ran3}). 

The topology of the constraint surface can be found in the shortest way, if we
consider ${\mathcal C}_1$ as a function ${\mathcal C}_1(B)$ of $B$ keeping all
other variables constant. Studying Eqs.\ (\ref{newcf}) and (\ref{dCdB}) we can
easily see that ${\mathcal C}_1(B)$ decreases in the interval $(-\infty, -{\sf
  M}^2/2)$ from $\infty$ to $9p^2$, it increases in the interval $( -{\sf
  M}^2/2), {\sf M}^2/2)$, from $9p^2$ to $9p^2 + 2{\sf M}^6$ and it decreases
again in the interval $({\sf M}^2/2, \infty)$ from $9p^2 + 2{\sf M}^6$ to
$-\infty$. It follows that there is only one solution for $p \neq 0$ (because
${\mathcal C}_1(B) = 9p^2$ for $B = {\sf M}^2$). As ${\mathcal C}_1({\sf M}^2)
= 9p^2 > 0$, the solution must satisfy the inequality $B > {\sf M}^2$. This
solution depends smoothly on the values of all other variables.  For $p = 0$,
there are, however, two solutions: one with $B = {\sf M}^2$, which is a
continuous extension of the previous $p \neq 0$ case, and a new one with $B =
-{\sf M}^2/2$, which appears when the local minimum at $-{\sf M}^2/2$ touches
the $B$-axis. This shows that the constraint surface consists of two
components: one is a seven-dimensional smooth manifold, let us denote it by
$\Gamma_7$, that lies in the region $B \geq {\sf M}^2$, and one is a
six-dimensional smooth manifold, let us denote it by $\Gamma_6$, that is
defined by Eqs.\ (\ref{static}).

Next, let us study the solvability of the constraint equation (\ref{newconstr})
with respect to the conserved momenta. The unique solution of the constraint
(\ref{newconstr}) with respect to $B$ at $\Gamma_7$ is given by 
\be 
  B = f(p,{\sf M}), 
\label{solB}
\ee
where
\be
  f(p,{\sf M}) := \frac{1}{2}\sqrt[3]{9p^2 + {\sf M}^6 + 3p\sqrt{9p^2 +
      2{\sf M}^6}} + 
      \frac{1}{2}\sqrt[3]{9p^2 + {\sf M}^6 - 3p\sqrt{9p^2 + 2{\sf M}^6}}.
\label{f}
\ee
The equation (\ref{newconstr}) has, in fact, three independent solutions, but
only one of them is real. For $p = 0$, $B$ reaches its minimum ${\sf M}^2$ on
$\Gamma_7$; thus, at $\Gamma_7$, 
\be
  B \geq {\sf M}^2.
\label{csolBG7}
\ee
Given $B$, one can solve uniquely for $\bar{E}$ and there is also a unique
solution for $[E]^2$. The solution for $\bar{E}$ is
\be
  \bar{E} = -\frac{1}{8x} - \frac{x[E]^2}{2} +\frac{x}{2}f(p,{\sf M}).
\label{solEG7}
\ee

Similarly, we can find the solution of the constraint (\ref{newconstr}) with
respect to ${\sf M}$. Consider ${\mathcal C}_1$ as a function ${\mathcal
  C}_1({\sf M})$ of ${\sf M}$ keeping all other variables fixed. ${\mathcal
  C}_1({\sf M})$ is symmetric with respect to the reflection on the ${\mathcal
  C}_1$-axis. We have to distinguish two cases: \underline{$B < 0,\,B = 0$}:
${\mathcal C}_1({\sf M})$ decreases in the interval $(-\infty, 2B)$ from
$\infty$ to $9p^2$, it increases in the interval $(2B,0)$ from $9p^2$ to the
local maximum $9p^2 - 4B^3$ at ${\sf M} = 0$, it decreases again in the
interval $(0,-2B)$ from $9p^2 - 4B^3$ to $9p^2$, and, finally, it increases in
$(-2B,\infty)$ from $9p^2$ to $\infty$. A solution exists only for $p = 0$,
and there are two solutions, then, ${\sf M} = \pm\sqrt{-2B}$. This is at the
surface $\Gamma_6$.  \underline{$B > 0$}: then, there is only one minimum, at
${\sf M} = 0$, with the value $9p^2 - 4B^3$; there are no other extrema.
Consider Eq.\ (\ref{newconstr}) as an equation for ${\sf M}^{-2}$. There are
three real solutions for ${\sf M}^{-2}$, but all of them are negative unless
\be
  4B^3 - 9p^2 \geq 0,
\label{csolMG7}
\ee
and then there is only one non-negative solution. A reasonable solution is,
therefore, unique, it depends continuously of the other variables and it lies
at $\Gamma_7$. We can write it in the form
\[
  {\sf M} = \pm\sqrt[4]{\frac{4B^3 -
  9p^2}{4B}}\left(\cos\frac{1}{3}\text{arctan}\frac{3p}{\sqrt{4B^3 -
  9p^2}}\right)^{-1/2}.
\]
This solution can, of course, be also expressed by means of square and third
roots, but only if one employs complex numbers. We can see that squaring of
$\sf M$ introduced non-physical solutions with negative $\sf M$; we have to
choose the positive branch. Observe that all points at the surface $\Gamma_7$
must satisfy the inequalities (\ref{csolBG7}) and (\ref{csolMG7}).

The equations of motion that follow from the action (\ref{newaction}) can be
written as follows
\bea
  \dot{\bar{E}} & = & [\dot{E}] = \dot{\sf M} = 0,
\label{dotmom} \\
  \dot{T}_2 & = & -\nu_1\frac{\partial {\mathcal C}_1}{\partial
  B}\frac{\partial B} {\partial \bar{E}},
\label{dotT2} \\
  \dot{T}_3 & = & -\nu_1\frac{\partial {\mathcal C}_1}{\partial
  B}\frac{\partial B} {\partial [E]},
\label{dotT3} \\
  \dot{\sf T} & = & \nu_1\frac{\partial {\mathcal C}_1}{\partial {\sf
  M}}, 
\label{dotT} \\
  \dot{p} & = & \nu_1\frac{\partial {\mathcal C}_1}{\partial
  B}\frac{\partial B} {\partial x},
\label{dotp} \\
  \dot{x} & = & -\nu_1\frac{\partial {\mathcal C}_1}{\partial p}.
  \label{dotx}
\eea
It is easy to derive the radial equation from them. On $\Gamma_6$, we have
only a static solutions:
\[
  \dot{x} = \dot{T} = \dot {p} = \dot{T}_2 = \dot{T}_3 = 0.
\]
On $\Gamma_7$, $\dot{\sf T} > 0$ for $\nu_1 > 0$ and Eqs.\ (\ref{dCdM}),
(\ref{dotT}) and (\ref{dotx}) yield
\[
  \frac{dx}{d{\sf T}} = -\frac{3p}{{\sf M}^3(2B + {\sf M}^2)}.
\]
Using the constraint (\ref{newconstr}) to exclude $p$, we obtain the radial
equation on $\Gamma_7$:
\be
  \frac{dx}{d{\sf T}} = \pm\frac{1}{{\sf M}^2}\sqrt{\frac{B}{{\sf M}^2} - 1}.
\label{radeqG7}
\ee
Let us compare it with Eq.\ (\ref{radeq}). Eqs.\ (\ref{dotmom}) and
(\ref{x}) imply that
\[
  \frac{dx}{d{\sf T}} = \frac{1}{{\sf M}^2}\frac{dR}{d{\sf T}},
\]
whereas Eqs.\ (\ref{B}), (\ref{x}) and (\ref{radpot}) give
\[
  V = 1 - \frac{B}{{\sf M}^2}.
\]
Hence, the radial equations (\ref{radeqG7}) and (\ref{radeq}) are equivalent,
and the dynamics of the new action on $\Gamma_7$ coincides with that of the
old action. $\Gamma_6$ consists of the unintentionally added unphysical
solutions, at least on the phase space defined by Eqs.\ (\ref{ran21}), but
nothing is added in the physical region.

Finally, we shall study the question of the monotonicity of the time functions
$T_2$, $T_3$ and $\sf T$ along the dynamical trajectories at $\Gamma_7$. Eqs.\
(\ref{dCdM}) and (\ref{dCdB}) show that the right-hand sides of Eqs.\
(\ref{dotT2}) and (\ref{dotT}) cannot vanish, and we have for $\nu_1 > 0$:
\[
 \dot{T}_2 > 0, \quad \dot{\sf T} > 0.
\]
Hence, $T_2$ and $\sf T$ are good times and, as we have seen above, the
constraint can be solved uniquelly for the corresponding momenta, $\bar{E}$
and $\sf M$. As for $T_3$, Eq.\ (\ref{dotT3}) implies that $\dot{T}_3 > 0$ for
$[E] > 0$, $\dot{T}_3 = 0$ for $[E] = 0$ and $\dot{T}_3 < 0$ for $[E] < 0$.
The solvability with respect to $[E]$ is also only partial: there is a unique
positive and a unique negative solution. Thus $T_3$ is a good time only in
some special cases, as the next section will show.

\section{Some interesting special cases}
\label{sec:cases}
In this section, we shall remove some degrees of freedom and describe the
resulting simpler models in terms of the new variables.

\subsection{Dust degrees of freedom removed}
Here, we remove the variables $\sf M$ and $\sf T$ and return to the system
considered in Ref.\ \cite{H1}. It has two degrees of freedom: the black hole
mass $E_-$ and the position $x$ of the shell. To this aim, we first choose a
particular value of $\sf M$ and demote it so to a mere parameter. In
this way, a submanifold $\Gamma_{7M}$ of $\Gamma_7$ emerges. Second, we
take a quotient of $\Gamma_{7M}$ by the $\sf T$-curves (see Ref.\ \cite{H2}
for more detail of this {\em exclusion of a cyclic variable}). The system that
results has the action
\[
  S_M = \int dt(-\bar{E}\dot{T}_2 - [E]\dot{T}_3 - p\dot{x} - \nu_1{\mathcal
  C}_1), 
\]
where ${\mathcal C}_1$ is given by Eq.\ (\ref{newcf}) as before, only $\sf M$
is a parameter now. The equations of motion comprise Eq.\ (\ref{dotmom})
without $\dot{\sf M} = 0$, as well as Eqs.\ (\ref{dotT2}), (\ref{dotT3}),
(\ref{dotp}) and (\ref{dotx}).

The properties of this system are analogous to that of the original one. In
particular, the constraint is regular everywhere on $\Gamma_{7M}$ because the
derivative (\ref{dCdp}) and (\ref{dCdB}) cannot both simultaneously vanish,
$T_2$ is a good time variable and the constraint equation is uniquely
solvable for $\bar{E}$.

\subsection{Black hole degree of freedom removed}
Here, we demote the variable $E_-$ to a parameter and remove the corresponding
cyclic variable $T_2/2 - T_3$; the remaining two degrees of freedom are the
internal energy $\sf M$ and the position $x$ of the shell. The system obtained
in this way will describe the dynamics of the dust shell in the field of a
fixed black hole if $E_- > 0$ or in the Minkowski space if $E_- = 0$. One
returns so to the system considered in Ref.\ \cite{Khaj}; the procedure, in
different coordinates, has been performed in Ref.\ \cite{H2}.

To begin with, we transform variables as follows:
\[
  \bar{E} = \frac{E_+ + E_-}{2},\quad T_2^+ = T_2/2 + T_3,
\]
\[
  [E] = E_+ - E_-,\quad T_2^- = T_2/2 - T_3.
\]
Then, we choose a particular value for $E_-$; this defines the black hole
mass, and formally, a six-dimensional submanifold $\Gamma_{7E}$ of
$\Gamma_7$. Finally, we take the quotient $\Gamma_{7E}/T_-$ of $\Gamma_{7E}$
by the $T_-$-curves. The result is the action
\[
  S_E = \int dt(-E_+\dot{T}_2^+ + {\sf M}\dot{\sf T} - p\dot{x} - \nu_1{\mathcal
  C}_2), 
\]
where
\[
  {\mathcal C}_2 = 9p^2 - (B_2 - {\sf M}^2)(2B_2 + {\sf M}^2),
\]
and
\[
  B_2 = \frac{1}{4x^2} +\frac{E_+ + E_-}{x} + (E_+ - E_-)^2.
\]

The variable $\sf T$ is a good time and $C_2 = 0$ is of course solvable for
$\sf M$ exactly as in the general case. The variable $T_2^+$ is, for large
values of the parameter $E_-$, not a good time, because it is a combination
$T_2/2 + T_3$ of $T_2$ that increases along all dynamical trajectories and
$T_3$ that increases along those with $E_+ > E_-$ and decreases along those
with $E_+ < E_-$. One can show that $\dot{T}_2^+$ is positive for $E_+ - E_- >
0$ and negative for $E_+ - E_- < -{\sf M}$ and that there is no surface for
$E_- > {\sf M}$ that would be transversal to both the dynamical trajectories
and the orbits of the continuous subgroup. \footnote{Still, one can transform
  $T_2^+$ so that the resulting time levels are transversal to the dynamical
  trajectories, but a mere shift is not sufficient; one must screw the level
  to a helix form with the helix axis in the $E_+$-direction; this may be done
  by a singular ``shift'' proportional to $(E_+-E_-)^{-1}$. We shall not go
  into detail.}  The derivative of ${\mathcal C}_2$ with respect to $E_+$
changes sign somewhere if $E_-$ is sufficiently large and so there are more
that one, or no solution of the constraint with respect to $E_+$.

There is one prominent exception: suppose that $E_- = 0$. Then, $E_+ -
E_-$ is positive if $E_+$ is. Hence, in the physical region (where $E_+ > 0$),
$T_2^+$ {\em is} a good time and the constraint has a {\em unique} solution
for $E_+$ there, namely
\[
  E_+ = -\frac{1}{2x} + \sqrt{f(p,{\sf M})},
\]
where $f$ is defined by Eq.\ (\ref{f}).

One can also remove all four variables $\sf T$, $\sf M$, $E_-$ and $T_2^-$ so
that only one degree of freedom, the position $x$ of the shell remains, but it
is not difficult to work out the properties of the system using the results
obtained in this section, and we left it as an easy exercise to the reader.

The special case $E_- = 0$ is also interesting, because it is a part of the
boundary of the physical region given by inequalities (\ref{ran3}) on
$\Gamma_7$. The variational principle (\ref{newaction}) does not break down at
the boundary, that is, it defines a regular dynamics there. From this, it must
clearly follow that the action defines a regular dynamics even at
those points of $\Gamma_7$ that do not satisfy the conditions (\ref{ran3}). We
have studied the meaning of this dynamics and of these points in Sec.\
\ref{sec:descr}.

\section{Acknowledgments}
Important discussions with Karel V. Kucha\v{r} are acknowledged. This work was
supported in part by the Swiss Nationalfonds, and by the Tomalla Foundation,
Zurich.


\begin{thebibliography}{99}
\bibitem{guth} E.~Farhi, A.~H.~Guth and J.~Guven, Nucl.\ Phys.\
  \textbf{B339}, 417 (1990).
\bibitem{HKK} P.~H\'{a}j\'{\i}\v{c}ek, B.~S.~Kay and K.~V.~Kucha\v{r},
  Phys.\ Rev.\ \textbf{D46}, 5439 (1992).
\bibitem{hooft} T.~Dray and G.~'t Hooft, Commun.\ Math.\ Phys.\
  \textbf{99}, 613 (1985).
\bibitem{KW} P.~Kraus and F.~Wilczek, Nucl.\ Phys.\ \textbf{B433}, 403
  (1995).
\bibitem{D} R.~Dautcourt, Math.\ Nachr.\ \textbf{27}, 277 (1964).
\bibitem{I} W.~Israel, Nuovo Cim.\ \textbf{44B}. 1 (1966);
  \textbf{48B}, 463 (1967).
\bibitem{berezin} V.~A.~Berezin, Phys.\ Lett.\ \textbf{B241},
  194 (1990).
\bibitem{HB} P.~H\'{a}j\'{\i}\v{c}ek and J.~Bi\v{c}\'{a}k, Phys.\
  Rev.\ \textbf{D56}, 4706 (1997).
\bibitem{louko} J.~L.~Friedman, J.~Louko, and S.~N.~Winters-Hilt,
  Phys.\ Rev.\ \textbf{D56}, 7674 (1997).
\bibitem{H3} P.~H\'{a}j\'{\i}\v{c}ek, Commun.\ Math.\ Phys.\ \textbf{150}, 545
  (1993). 
\bibitem{Hkij} P.~H\'{a}j\'{\i}\v{c}ek and J.~Kijowski, Phys.\ Rev.\
  \textbf{D57}, 914 (1998).
\bibitem{ansold} A.~Ansoldi, A.~Aurilia, R.~Balbinot and E.~Spalluci, Class.\
  Quant.\ Grav.\ \textbf{14}, 2727 (1997).
\bibitem{Khaj} K.~V.~Kucha\v{r}, {\em Time for space: canonical dynamics of
  gravitating shells}, in preparation. 
\bibitem{H1} P.~H\'{a}j\'{\i}\v{c}ek, Phys.\ Rev.\ \textbf{D57}, 936 (1998).
\bibitem{H2} P.~H\'{a}j\'{\i}\v{c}ek, {\em Relation between the guessed and the
  derived super-Hamiltonian for the spherically symmetric shells}, Preprint,
  gr-qc/9804010.
\bibitem{nucl} P.~H\'{a}j\'{\i}\v{c}ek, Nucl.\ Phys.\ \textbf{B} (Proc.\
  Suppl.) \textbf{57}, 115 (1997).
\bibitem{HK1} P.~H\'{a}j\'{\i}\v{c}ek and K.~V.~Kucha\v{r}, Phys.\ Rev.\
  \textbf{D41}, 1091 (1990).
\bibitem{HK2} P.~H\'{a}j\'{\i}\v{c}ek and K.~V.~Kucha\v{r}, J. Math.\ Phys.\
  \textbf{31}, 1723 (1990).
\bibitem{souriau} J.-M.~Souriau, {\em Structure des syst\`{e}mes
    dynamiques}. Paris, Dunod, 1970.


\end{thebibliography}
\end{document}